\documentclass[%
reprint,
superscriptaddress,
 amsmath,amssymb,
 aps,
]{revtex4-1}

\usepackage{graphicx}
\usepackage{dcolumn}
\usepackage{bm}
\usepackage[mathlines]{lineno}
\usepackage[colorlinks=true]{hyperref}
\usepackage{mathrsfs}
\usepackage{xcolor}
\usepackage{float}
\usepackage{ulem}
\usepackage[utf8]{inputenc}

\newcommand{\be}{\begin{equation}}
\newcommand{\ee}{\end{equation}}
\newcommand{\bal}{\begin{aligned}}
\newcommand{\eal}{\end{aligned}}
\newcommand{\bea}{\setlength\arraycolsep{2pt} \begin{eqnarray}}
\newcommand{\eea}{\end{eqnarray}}

\begin{document}


\title{Relaxation rate of RNdS black hole}
\author{Ming Zhang}
\email{mingzhang@jxnu.edu.cn}
\affiliation{Department of Physics, Jiangxi Normal University, Nanchang 330022, China}
\author{Jie Jiang}
\email{jiejiang@mail.bnu.edu.cn}
\affiliation{Department of Physics, Beijing Normal University, Beijing 100875, China}
\author{Zhen Zhong}
\email{Corresponding author: zhenzhong@mail.bnu.edu.cn}
\affiliation{Department of Physics, Beijing Normal University, Beijing 100875, China}

\date{\today}

\begin{abstract}
We investigate the relaxation rate of the RNdS black hole perturbed by neutral massless scalar field in the eikonal limit. We find that the fastest relaxation rate of the composed system increases with the cosmological constant for all spacetime dimensions. We also find that, when the  cosmological constant decreases from maximum value to zero, the corresponding critical charge of the four-dimensional RNdS black hole maximizing the relaxation rate of the composed system gradually decreases monotonically to $\bar{Q}=0.726$. However, for the higher dimensional systems, this is not the case and the critical charges decrease from maximum to zero.
\end{abstract}


\maketitle


\section{Introduction}
Quasinormal modes (QNMs) is the damped oscillations of the black hole/field system that is impinged upon by perturbations \cite{Berti:2009kk}. At the end stage of the black hole merger, there will be gravitational waves in form of QNMs. Investigations of the QNM was started more than five decades ago and the recent interest of it is its application on the Penrose strong cosmic censorship conjecture \cite{Cardoso:2017soq,Hod:2018dpx} and on the black hole's no-hair theorem \cite{Hod:2015hza,Hod:2016bas,Hod:2017www,Hod:2018fet,Hod:2017kpt,Huang:2017whw,Li:2019tns}.  As linear perturbations, the QNMs can usually be described by the form $e^{-i\omega t}$, where $t$ is the time and $\omega$ is the complex quasinormal resonance spectrum function \cite{Nollert:1999ji,Cardoso:2001hn,Hod:1998vk}. The imaginary part of $\omega$ contributes to the damping while the real part sees the oscillation of the perturbation. Physically, the spectrum function should be purely ingoing at the horizon of the black hole and it should be purely outgoing at spatial infinity for asymptotically flat or de Sitter (dS) spacetime. (For the Anti-de Sitter (AdS) spacetime, it should be finite at spatial infinity) \cite{Berti:2009kk}. As a result, the resonance spectrum is discrete and $\omega=\omega_n$, with $n$ the overtone number. If $\text{Im}(\omega)>0$, the QNMs denote an unstable perturbation. 

The uniqueness theorem \cite{Carter:1971zc, Hawking:1971vc, Robinson:1975bv} establishes that there will be at most three conserved quantities, mass, electric charge and angular momentum, for the Kerr-Newman black hole family in Einstein-Maxwell theory. Based on it, the no-hair conjecture \cite{Ruffini:1971bza} confirms that perturbations of matter fields on black holes in the Einstein-Maxwell family will finally die out as time elapses. Correspondingly, the QNMs as a description of the interaction between black hole and exterior matter field will be dominated by the damping resonance spectrum. However, it has recently been numerically shown \cite{Zhu:2014sya,Konoplya:2014lha} and analytically solved \cite{Hod:2018fet} that the Reissner–Nordström (RN) black hole in the asymptotically dS spacetime will be unstable with perturbation originated from charged scalar fields. 

The relaxation time, which is related to the fundamental quasinormal modes by 
\begin{equation}
\tau=\omega_I^{-1}(n=0),
\end{equation}
is a characteristic quantity reflecting the relaxation rate of the perturbation impinging upon the black hole by the matter field. The relaxation time of the RN black hole perturbed by charged massive scalar field was studied in \cite{Hod:2016jqt,Zhang:2018jgj} and  the relaxation rate for the composed RN black hole/massless scalar field system was investigated in  \cite{Hod:2018ifz,Zhang:2019ynp,Pani:2013wsa,Pani:2013ija}. In the present paper, we will explore the relaxation time of the RNdS black hole with neutral massless scalar field perturbation. We will write down the analytical quasinormal resonance frequency for the system in the eikonal regime and then show the effects of the positive cosmological constant on the relaxation rate of the composed system.

We will arrange the remaining part of the paper as follows. In section \ref{chapter2}, we will review an ordinary differential equation describing the RNdS black hole perturbed by the neutral massless scalar field. In section \ref{chapter3}, we will  show the quasinormal resonance frequency of the system composed of RNdS black hole and neutral massless scalar field and analyze the relaxation rate of it. Section \ref{chapter4} will be devoted to our conclusion and discussion.

\section{Description of RN\lowercase{d}S black hole perturbed by neutral massless scalar field}\label{chapter2}
The line element of the RNdS black hole can be described in the Schwarzschild-like coordinates as
\begin{equation}
ds^2=-f(r)dt^2+\frac{dr^2}{f(r)}+r^2 d\Omega^2_{d-2},
\end{equation}
with
\begin{equation}
f(r)=1-\frac{16 \pi  M r^{3-d}}{(d-2) \Omega_{d-2} }+\frac{32 \pi ^2 Q^2 r^{-2 (d-3)}}{\left(d^2-5 d+6\right) \Omega_{d-2} ^2}-\frac{r^2}{L^2}.
\end{equation}
$M,\,Q$ are individually the mass and electric charge of the RNdS black hole. $L$ is the dS radius and it is related to the positive cosmological constant $\Lambda$ by $L^2=(d-1)(d-2)/(2\Lambda)$. $\Omega_{d-2}$ is the volume of the $(d-2)$-sphere. There are three horizons for the black hole, namely the innermost Cauchy horizon $r_-$, the event horizon $r_+$ and the outmost cosmological horizon $r_c$, which are solutions of the equation $f(r)=0$.

The propagation function of the neutral massless scalar field $\Psi (t,r,\Theta)$  in the RNdS spacetime can be described by the Klein-Gordon wave equation
\begin{equation}\label{kg}
\triangledown^{a}\triangledown_{a}\Psi(t,r,\Theta)=0.
\end{equation}
As the spacetime background we consider here is spherically symmetric and the scalar perturbation is linear, we can expand the field $\Psi$ in terms of minimally coupled spherical harmonics
\begin{equation}
\sum_{lm}e^{-i\omega t}\frac{R_{lm}(r, \omega)}{r^{\frac{d-2}{2}}}Y_{lm}(\theta),
\end{equation}
where $l$ is the spherical harmonic index (angular number) and $m$ is the azimuthal harmonic index. We here will not show the equation of angular components $Y_{lm}$. The radial dynamics of the neutral massless scalar field perturbation of the RNdS black hole can be described by the differential equation
\begin{equation}\label{radial}
 f(r)^{2}\frac{d^2R(r)}{dr^2} +\frac{df(r)}{dr}f(r) \frac{dR(r)}{dr}+U R(r)=0,
\end{equation}
where
\begin{equation}
U=\omega^2-\frac{(d-4) (d-2) f(r)^2}{4 r^2} -\frac{f(r) \left[(d-2) r f'(r)+2 K_l\right]}{2 r^2},
\end{equation}
with $K_l=l(d-3+l)$. After introducing the tortoise coordinate $dr_{*}=dr/f(r)$, the radial perturbation equation can further be written in a Schr\"{o}dinger-like form,
\begin{equation}\label{sch}
\frac{\text{d}^2 R}{\text{d}r_{*}^2}+(\omega^2 -V)R=0,
\end{equation}
where the radial effective potential can be read off
\begin{equation}\label{pote}
 V=f(r) H(r),
\end{equation}
with
\begin{equation}
\begin{aligned}
H(r)=&\frac{K_l}{r^2}-\frac{(d-2) d r^2}{4 L^2}+\frac{d^2-6 d+8}{4 r^2}\\&+\frac{4 \pi  (d-2) M}{\Omega_{d-2}  r^{d-1}}-\frac{8 \pi ^2 (3 d-8) Q^2}{(d-3) \Omega_{d-2} ^2 r^{2 d-4}}.
\end{aligned}
\end{equation}

To single out a set of discrete quasinormal resonance frequency, we should impose the boundary conditions
\begin{equation}
R(r\to r_+)\sim e^{-i\omega r_{*}},\quad R(r\to r_c)\sim e^{i\omega r_{*}}
\end{equation}
on the system. Then we can calculate the resonant modes $\omega_{n}$ describing the relaxation dynamics of the RNdS black hole perturbed by the neutral massless scalar field.

\section{ Quasinormal modes of RN\lowercase{d}S black hole against neutral massless scalar field}\label{chapter3}
For the event horizon radius $r_+$ of the RNdS black hole, we have
\begin{equation}
\begin{aligned}
f(r_+)=0<1-\frac{16 \pi  M r_+^{3-d}}{(d-2) \Omega_{d-2} }+\frac{32 \pi ^2 Q^2 r_+^{-2 (d-3)}}{\left(d^2-5 d+6\right) \Omega_{d-2} ^2},
\end{aligned}
\end{equation}
which gives
\begin{equation}\label{evcon}
\bal
r_+>&\left(\frac{4 \pi}{\Omega_{d-2}} \right)^{\frac{1}{d-3}}\\&\times\left(\frac{2 M}{d-2}+\sqrt{\frac{4 M^2}{(d-2)^2}-\frac{2 Q^2}{d^2-5 d+6}}\right)^{\frac{1}{d-3}}.
\eal
\end{equation}
For a spatial position $r>r_+$ outside the black hole, we have
\begin{equation}
\frac{4 \pi  (d-2) M}{\Omega_{d-2}  r^{d-3}}<\frac{(d-2)^{2}}{2},
\end{equation}
\begin{equation}
\frac{8 \pi ^2 (3 d-8) Q^2}{(d-3) \Omega_{d-2} ^2 r^{2 d-6}}<\frac{(d-2)(3d-8)}{4}.
\end{equation}
Then in the eikonal regime $l\gg 1$,
we further have
\begin{equation}\begin{aligned}
&H(r)=\frac{K_{l}}{r^{2}} \left[1+\frac{d^2-6 d+8}{4 K_{l}}\right.\\&~~+\left.\frac{4 \pi  (d-2) M}{\Omega_{d-2}  r^{d-3}K_{l}}-\frac{8 \pi ^2 (3 d-8) Q^2}{(d-3) \Omega_{d-2} ^2 r^{2 d-6}K_{l}}-\frac{(d-2) d r^2}{4K_l L^2}\right]\\&\quad\quad\sim \frac{K_{l}}{r^{2}},
\end{aligned}\end{equation}
where we have used the condition
\begin{equation}
r<r_c<L.
\end{equation}

As a result, we obtain the approximated radial effective potential of the composed RNdS black hole/neutral massless scalar field system
\begin{equation}\label{appveff}
V=\frac{K_{l}f(r)}{r^{2}},
\end{equation}
which plays as a role of an effective potential barrier, and its maximum value locates at
\begin{equation}
r_0=\left[\frac{4 \pi  (d-1) M}{(d-2) \Omega_{d-2}}\left[1+\sqrt{1-\frac{2 (d-2)^2 Q^2}{(d-3) (d-1)^2 M^2}}\right]\right]^{\frac{1}{d-3}}.
\end{equation}

The WKB method is a semianalytic technique to which we can resort to obtain the complex quasinormal resonance frequency of the system composed by the RNdS black hole and the neutral massless scalar field and characterized by the approximated effective potential (\ref{appveff}) in the eikonal regime. The WKB resonance equation is \cite{Konoplya:2003ii,Iyer:1986np,Schutz:1985zz}
\begin{equation}\label{wkb}
  \frac{\mathcal{Q}_0}{\sqrt{-2\mathcal{Q}_0^{(2)}}}=-i\left(n+\frac{1}{2}\right)+\mathcal{O}\left(\frac{1}{l}\right),
\end{equation}
where we have denoted
\begin{equation}
\mathcal{Q}_{0}\equiv\omega^2 -V(r=r_{0}),
\end{equation}
\begin{equation}
\mathcal{Q}_0^{(2)}\equiv \left.\frac{\text{d}^2 \mathcal{Q}[r(r_{*})]}{\text{d}r_{*}^{2}}\right |_{r=r_{0}}.
\end{equation}
Explicitly, the WKB equation can be written as
\begin{equation}\label{reso}
 \frac{\omega^2-\frac{K_{l}f(r_{0})}{r_{0}^{2}}}{f(r_{0})\sqrt{-2V_{0}^{\prime\prime}}}=-i(n+\frac{1}{2}),
\end{equation}
where we denote $V_0=V(r=r_0)$. This equation is obtained in the condition of eikonal regime, where the spherical harmonic index $l$ is large. To further decouple the complex equation, we here impose a strong inequality between the imaginary part and the real part of the quasinormal resonance frequency, 
\begin{equation}\label{rbiggeri}
\omega_I\ll \omega_R.
\end{equation}

\begin{figure}[!htbp] 
   \centering
   \includegraphics[width=3.2in]{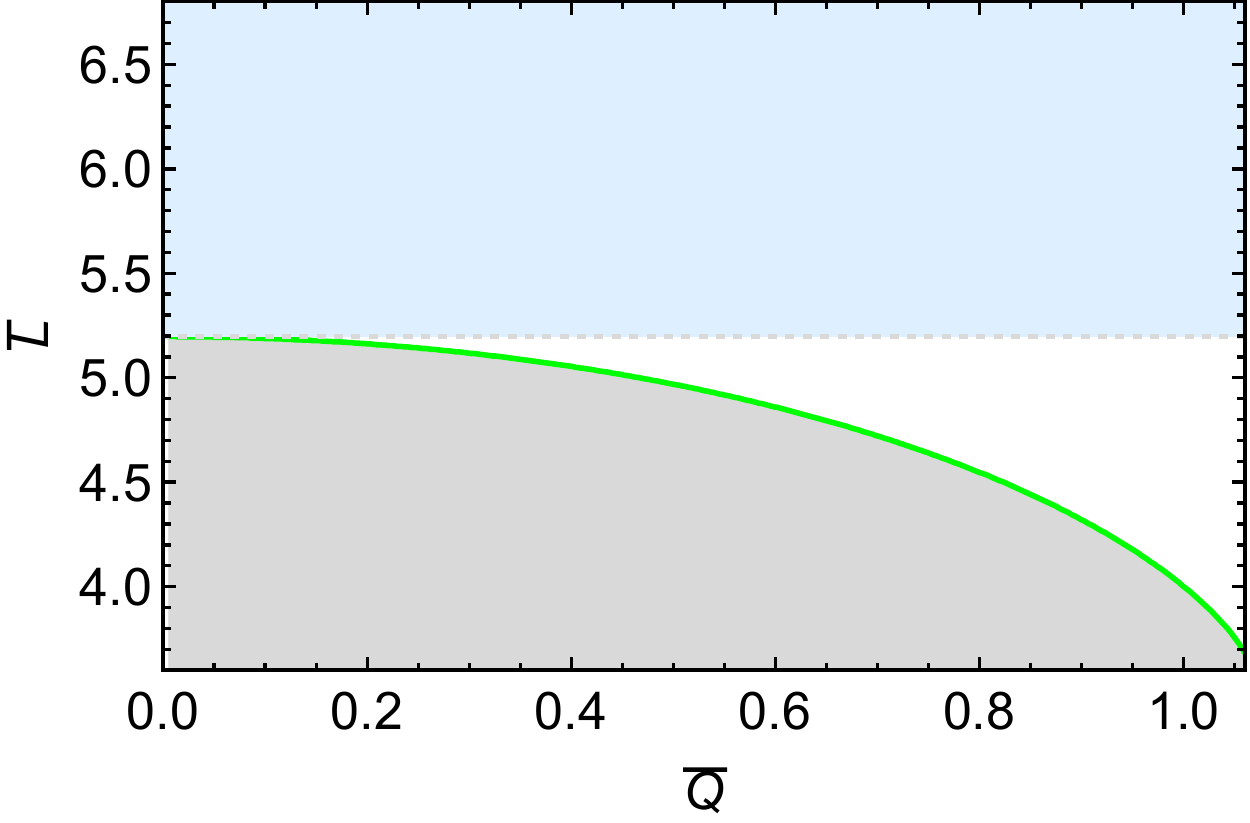}\\ \includegraphics[width=3.2in]{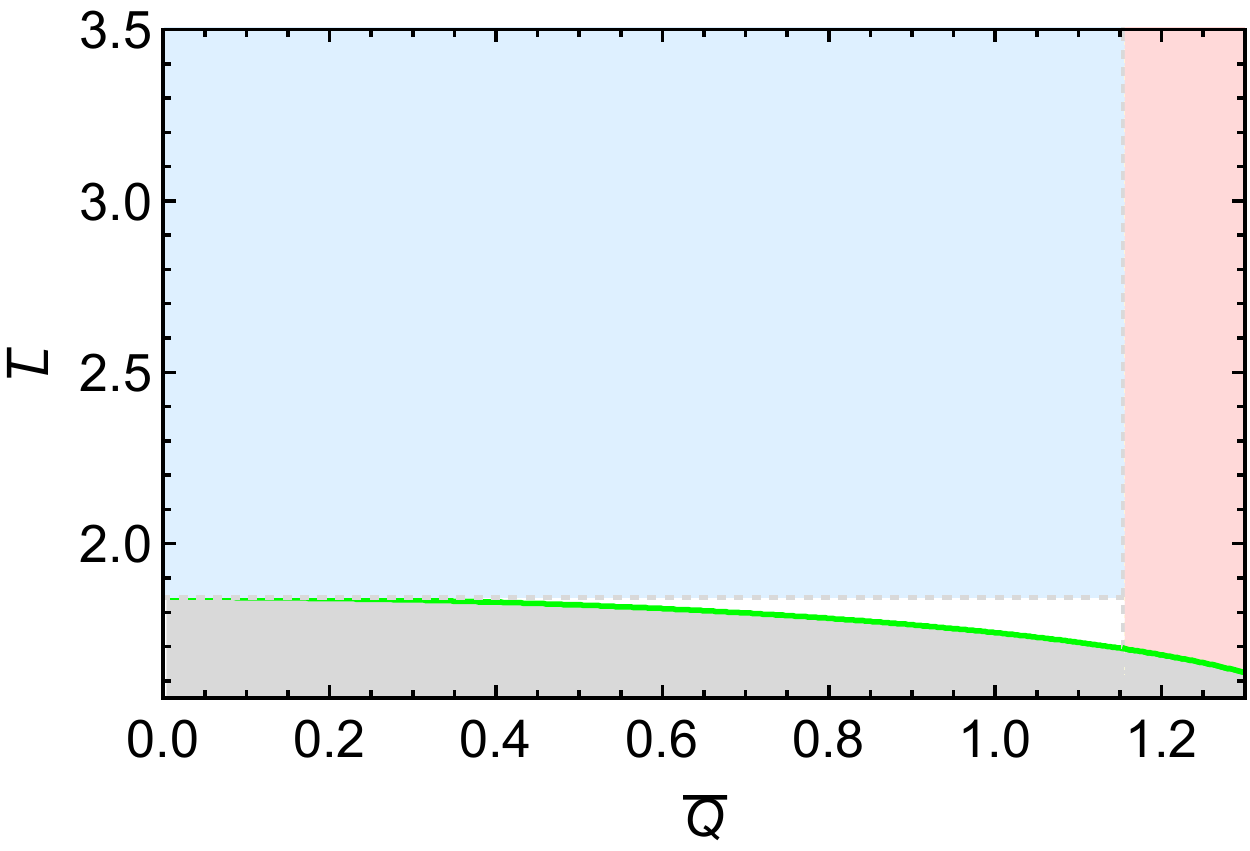}\\ \includegraphics[width=3.2in]{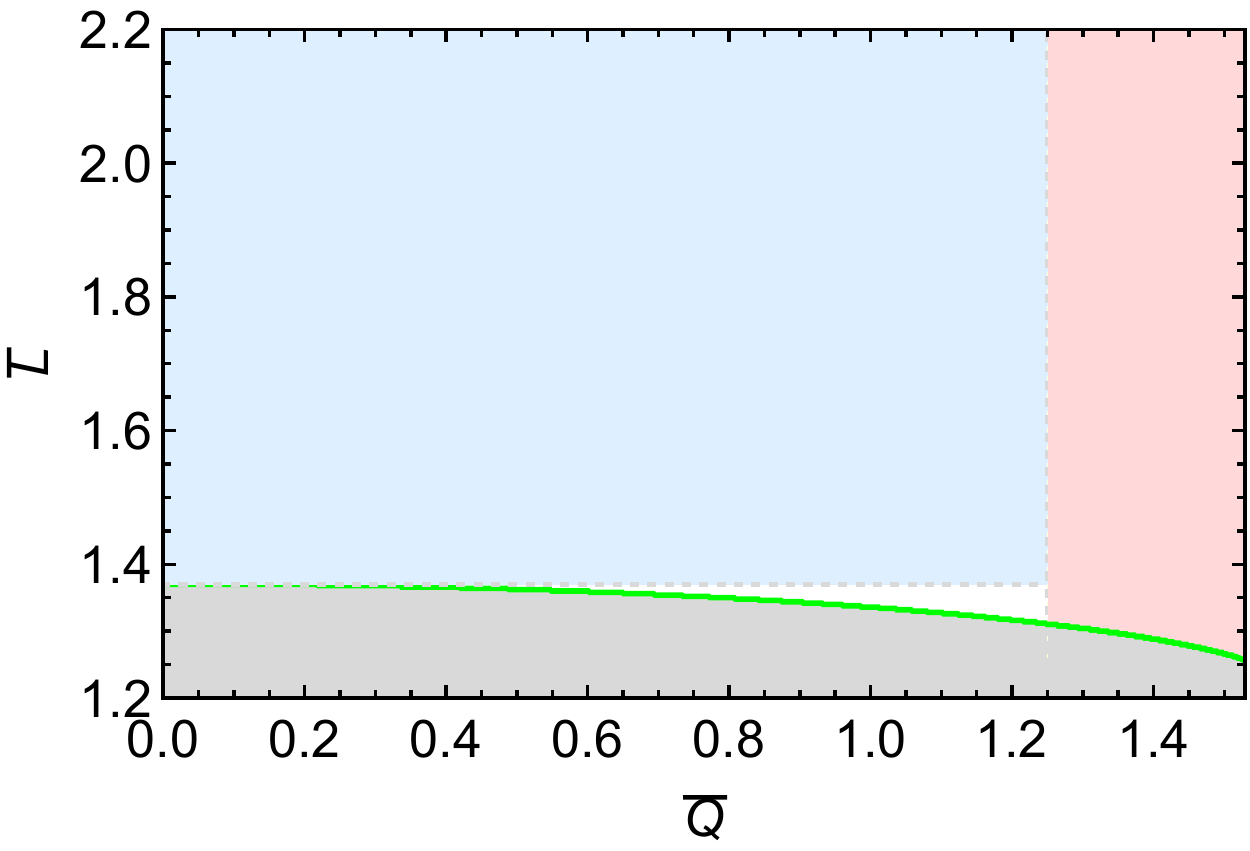}
   \caption{The parameter space of the composed four-dimensional, five-dimensional and six-dimensional composed RNdS black hole/neutral massless scalar field systems with $M=1$. In the parameter region beyond the green line (including it), the black hole has three horizons (the Cauchy horizon, event horizon and cosmological horizon). The upper diagram (four-dimensional case):  The gray region is forbidden; in the white region including the green line, the system has quasinormal resonance frequency whose corresponding relaxation rate is of the type that decreases first and then increases  in certain range of $\bar{Q}$; in the blue region, the system has quasinormal resonance frequency whose corresponding relaxation time is of the type that decreases first and then increases for $0\leqslant Q<Q_{\text{ext}}$ with $Q_{\text{ext}}$ the maximum permitted value of the electric charge of the RNdS black hole, see App. \ref{ap} for the four-dimensional case. The middle diagram (five-dimensional case) and the bottom diagram (six-dimensional case): The red regions are also forbidden as the requests of Eqs. (\ref{req1}), (\ref{req2}).   
For the three diagrams, the blue regions intersect with the gray regions at $\bar{L}=5.12, 1.84, 1.37$ $(\bar{\Lambda}=0.11, 1.77,  5.33)$, respectively.}
   \label{ps2311}
\end{figure}

\begin{figure*}[!htbp] 
   \centering
   \includegraphics[width=3.45in]{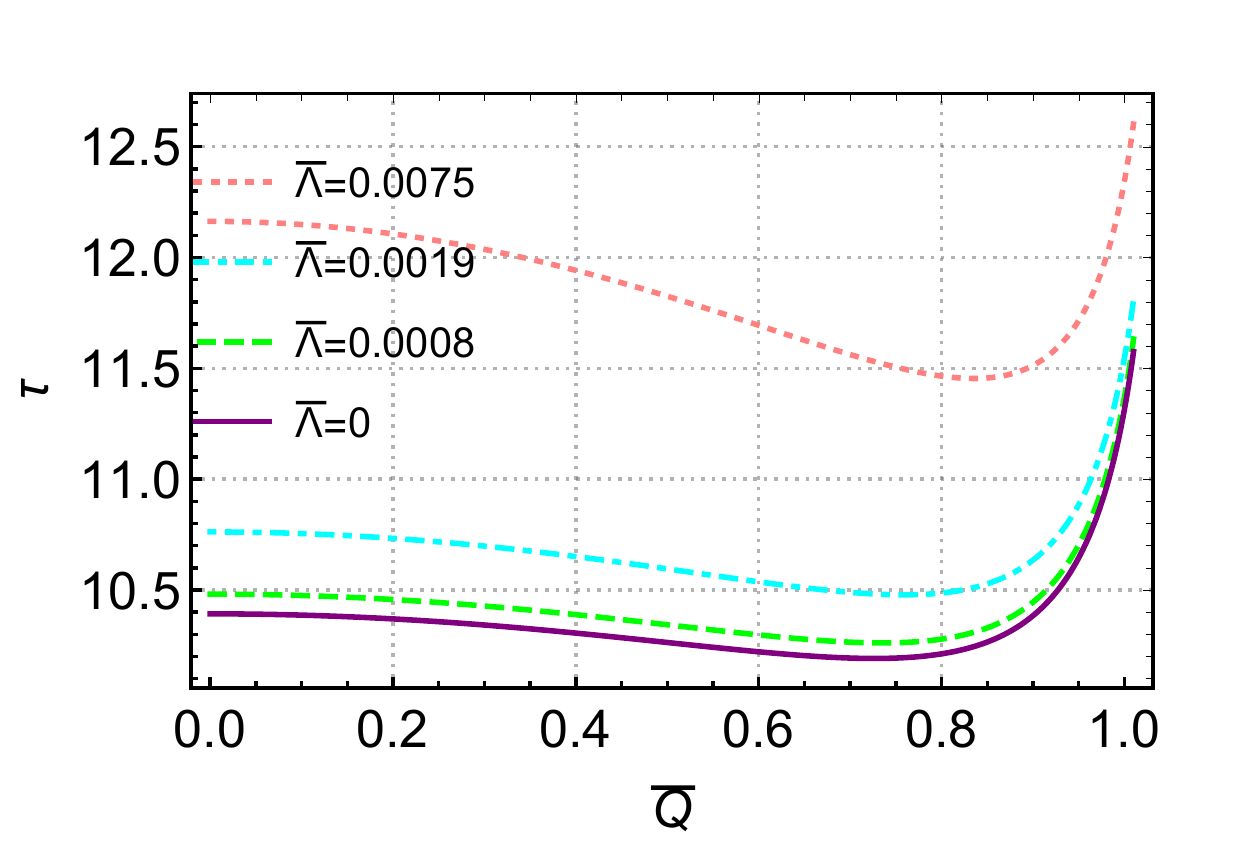}
    \includegraphics[width=3.35in]{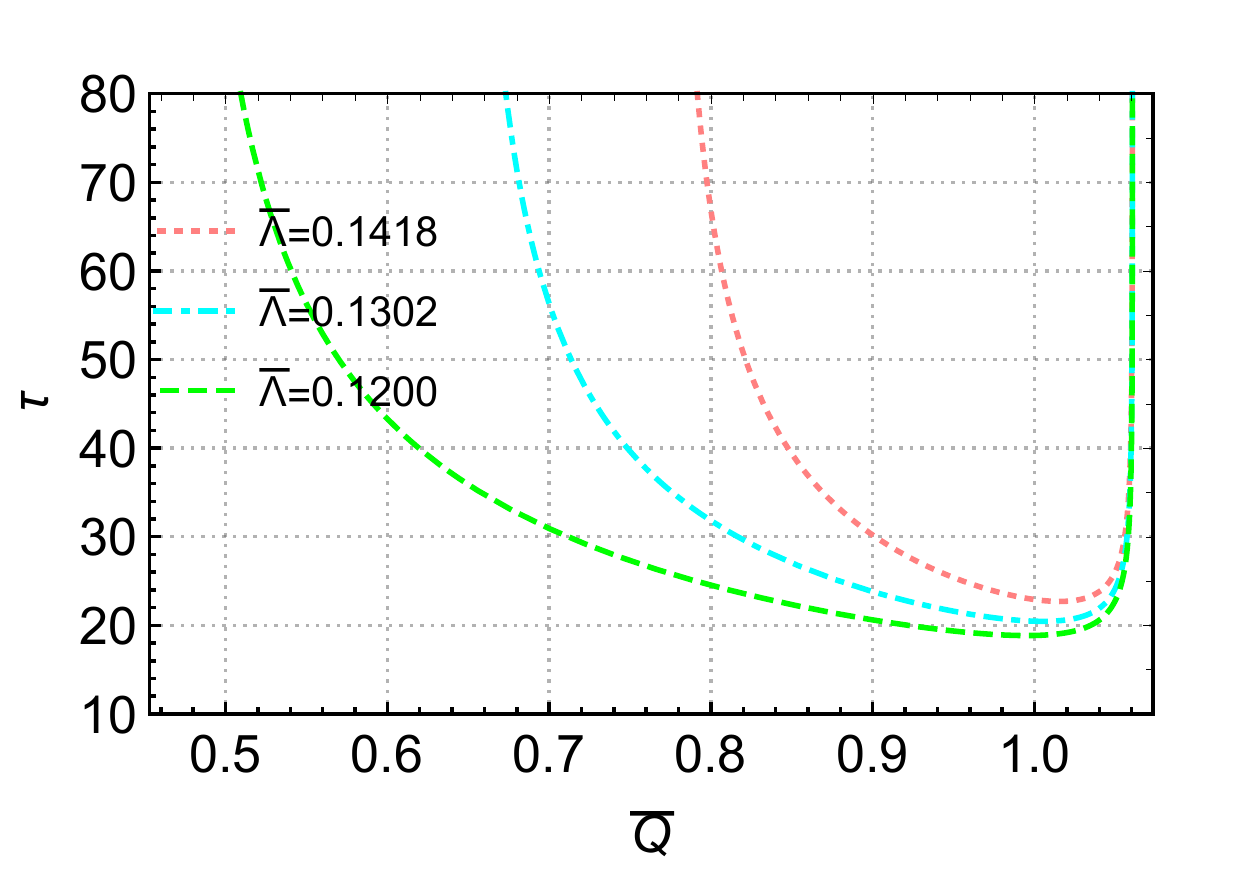}
   \caption{The relaxation rate of the four-dimensional RNdS black hole against the neutral massless scalar field perturbation with $M=1$. The left diagram corresponds to the blue region and the right diagram corresponds to the white region in the upper diagram of Fig. \ref{ps2311}.}
   \label{ps1911}
\end{figure*}

\begin{figure*}[!htbp] 
   \centering
    \includegraphics[width=3.45in]{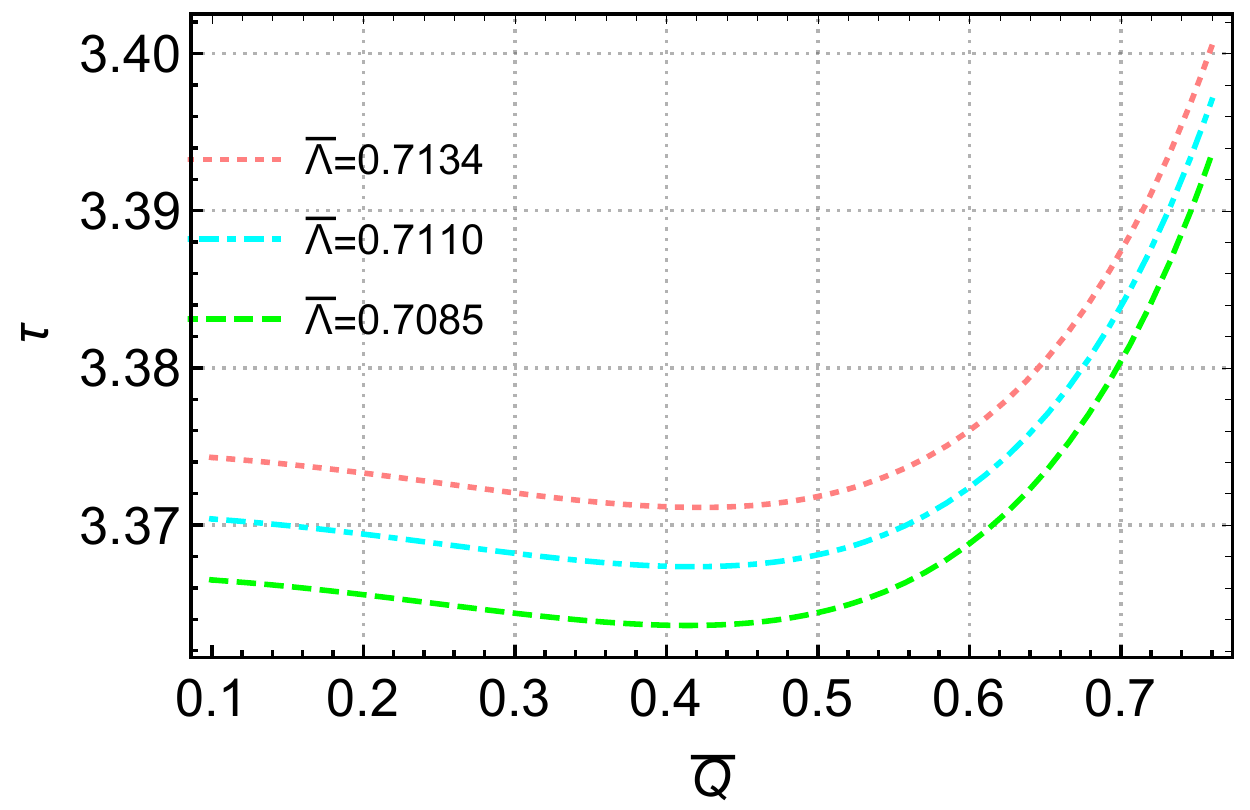}
     \includegraphics[width=3.35in]{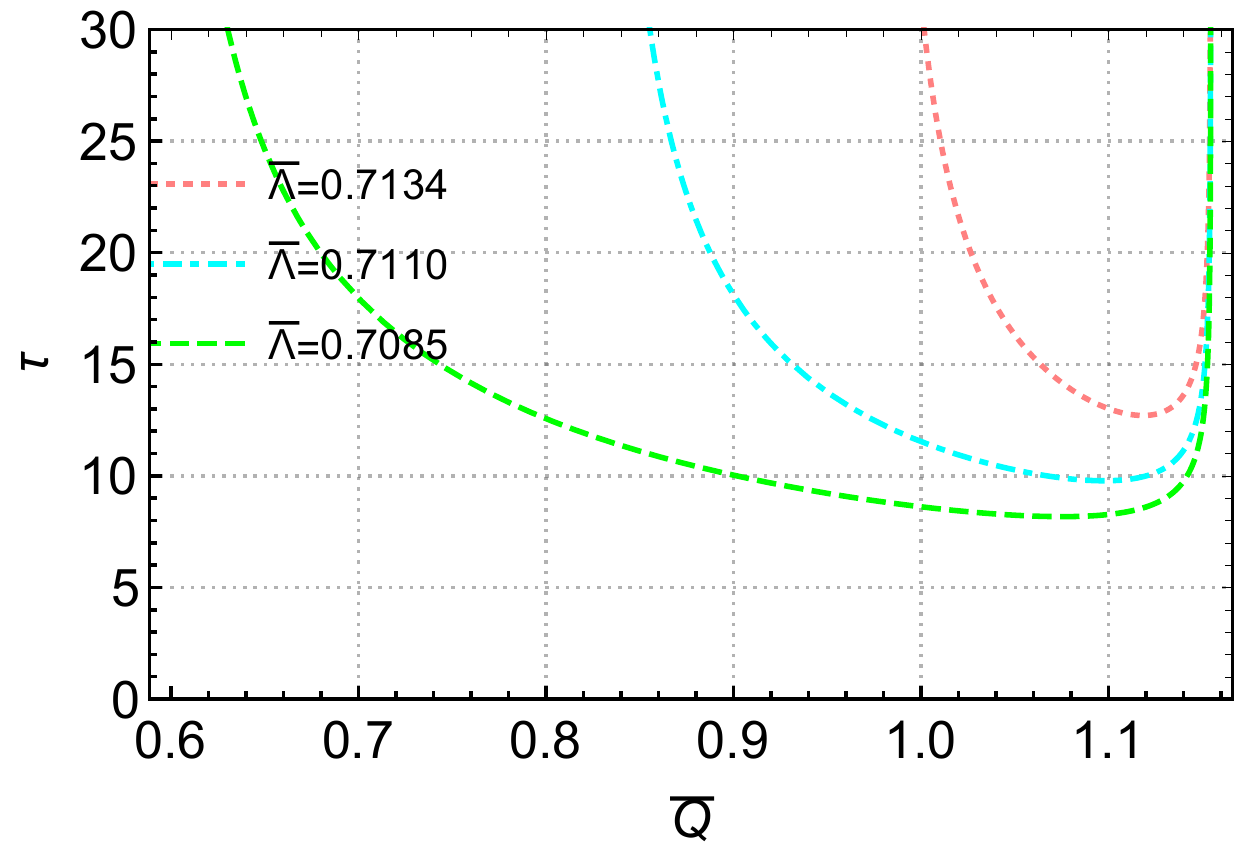}
   \caption{The relaxation time of five-dimensional RNdS black hole against neutral massless scalar field perturbation with $M=1$. The left diagram corresponds to the blue region and the right diagram corresponds to the white region in the middle diagram of Fig. \ref{ps2311}.}
   \label{rt1816}
\end{figure*}

\begin{figure*}[!htbp] 
   \centering
    \includegraphics[width=3.25in]{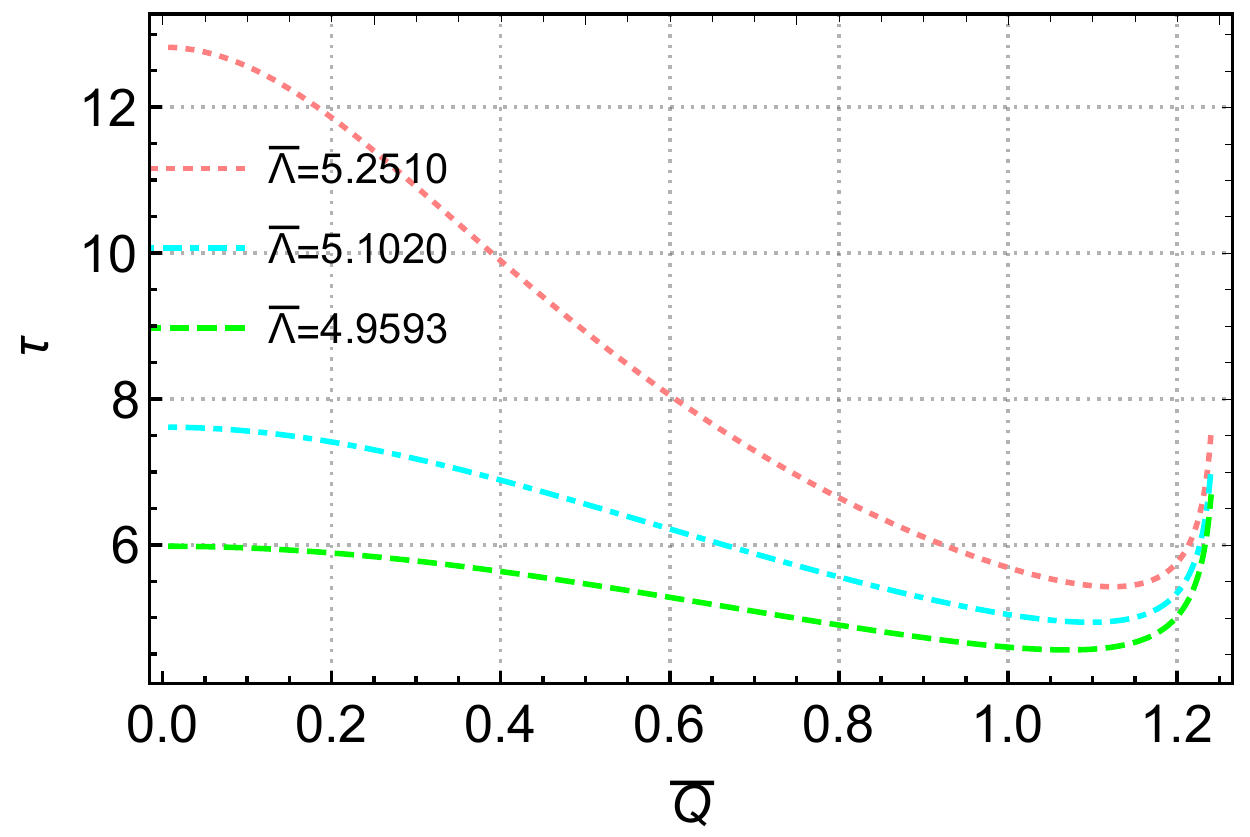}
     \includegraphics[width=3.25in]{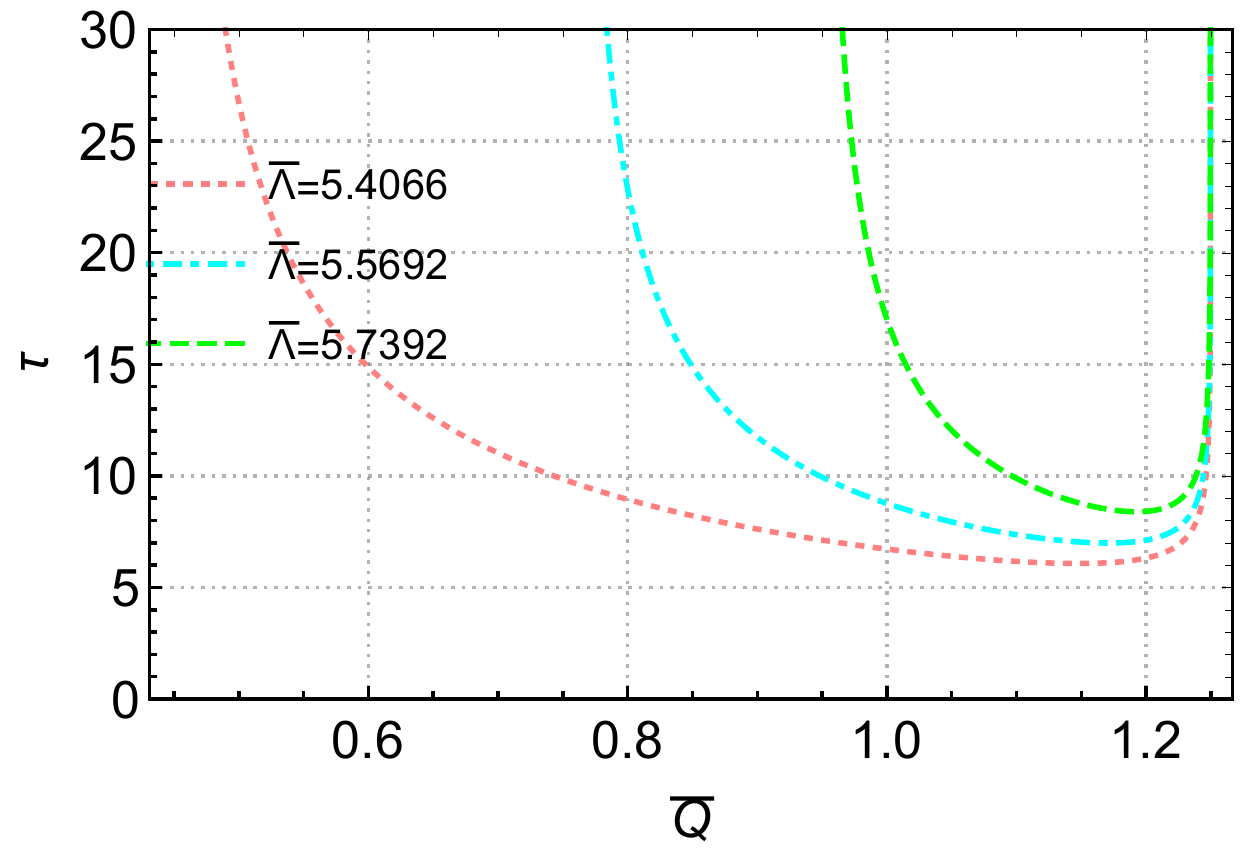}
   \caption{The relaxation time of six-dimensional RNdS black hole against neutral massless scalar field perturbation with $M=1$. The left diagram corresponds to the blue region, the right diagram corresponds to the white region in the bottom diagram of Fig. \ref{ps2311}.}
   \label{rt1618}
\end{figure*}

\begin{table*}[!htbp]
\centering
\caption{The fastest relaxation rate of the composed four-dimensional RNdS black hole/neutral massless scalar field system for $M=1,~n=0$.}
\begin{tabular}{cccccccc} 
  \hline\hline
$\Lambda~~~$ &~~~0~~~&~~~0.002~~~&~~~0.008~~~&~~~0.030~~~&~~~0.120~~~&~~~0.130~~~&~~~0.142\\
$\tau_{\text{min}}~~~$ &~~~10.191~~~&~~~11.522~~~&~~~11.497~~~&~~~11.454~~~&~~~18.846~~~&~~~20.453~~~&~~~22.698\\
$\bar{Q}_e~~~$ &~~~0.726~~~&~~~0.735~~~&~~~0.759~~~&~~~0.835~~~&~~~0.995~~~&~~~1.005~~~&~~~1.015\\
\hline
\hline
\end{tabular}
  \label{tbl:table1}
\end{table*}

\begin{table*}[!htb]
\centering
\caption{The fastest relaxation rate of the composed five-dimensional RNdS black hole/neutral massless scalar field system for $M=1,~n=0$.}
\begin{tabular}{ccccccccccc} 
  \hline\hline
$\Lambda~~~$ &~~~0~~~&~~~0.590~~~&~~~0.593~~~&~~~0.601~~~&~~~0.709~~~&~~~0.711~~~&~~~0.713~~~&~~~1.852~~~&~~~1.959~~~&~~~2.076\\
$\tau_{\text{min}}~~~$ &~~~2.606~~~&~~~3.192~~~&~~~3.197~~~&~~~3.208~~~&~~~3.364~~~&~~~3.367~~~&~~~3.371~~~&~~~8.174~~~&~~~9.780~~~&~~~12.706\\
$\bar{Q}_e~~~$ &~~~0~~~&~~~0.029~~~&~~~0.080~~~&~~~0.133~~~&~~~0.414~~~&~~~0.418~~~&~~~0.422~~~&~~~1.075~~~&~~~1.098~~~&~~~1.119\\
\hline
\hline
\end{tabular}
  \label{tbl:table2}
\end{table*}
\begin{table*}[!htb]
\centering
\caption{The fastest relaxation rate of the composed six-dimensional RNdS black hole/neutral massless scalar field system for $M=1,~n=0$.}
\begin{tabular}{ccccccccccc} 
  \hline\hline
$\Lambda~~~$ &~~~0~~~&~~~3.114~~~&~~~3.121~~~&~~~3.156~~~&~~~4.956~~~&~~~5.102~~~&~~~5.251~~~&~~~5.407~~~&~~~5.569~~~&~~~5.739\\
$\tau_{\text{min}}~~~$ &~~~1.581~~~&~~~2.452~~~&~~~2.456~~~&~~~2.475~~~&~~~4.559~~~&~~~4.938~~~&~~~5.425~~~&~~~6.075~~~&~~~6.991~~~&~~~8.393\\
$\bar{Q}_e~~~$ &~~~0~~~&~~~0.056~~~&~~~0.095~~~&~~~0.196~~~&~~~1.0690~~~&~~~1.0970~~~&~~~1.1220~~~&~~~1.1470~~~&~~~1.1700~~~&~~~1.1920\\

\hline
\hline
\end{tabular}
  \label{tbl:table3}
\end{table*}

\begin{table*}[]
\centering
\caption{Comparison of the fastest relaxation rate obtained between WKB method and a spectral method for $M=1,~n=0$.}
\begin{tabular}{|c|c|c|c|c|c|c|}
\hline
       & \multicolumn{2}{c|}{$d=4$}                                                                                         & \multicolumn{2}{c|}{$d=5$}                                                                                              & \multicolumn{2}{c|}{$d=6$}                                                                                            \\ \hline
       & \begin{tabular}[c]{@{}c@{}}$\bar{L}=40$\\$\bar{Q}=0.735$\end{tabular} & \begin{tabular}[c]{@{}c@{}}$\bar{L}=4.8$\\ $\bar{Q}=1.005$\end{tabular} & \begin{tabular}[c]{@{}c@{}}$\bar{L}=3.19$\\ $\bar{Q}=0.0239$\end{tabular} & \begin{tabular}[c]{@{}c@{}}$\bar{L}=1.75$\\ $\bar{Q}=1.0976$\end{tabular} & \begin{tabular}[c]{@{}c@{}}$\bar{L}=1.792$\\ $\bar{Q}=0.056$\end{tabular} & \begin{tabular}[c]{@{}c@{}}$\bar{L}=1.34$\\ $\bar{Q}=1.17$\end{tabular} \\ \hline
$\tau_{\mathrm{min}}^{\mathrm{WKB}}(l\to\infty)$ & 10.2612                                                & 20.4527                                                 & 3.1923                                                    & 9.7803                                                    & 2.4518                                                    & 6.9911                                                  \\ \hline
$\tau_{\mathrm{min}}^{\mathrm{spectral}}(l=30)$      & 10.2608                                               & 20.4510                                                 & 3.1917                                                    & 9.7803                                                   & 2.4513                                                    & 6.9915                                                  \\ \hline
\end{tabular}
  \label{tbl:table4}
\end{table*}

Consequently, we can decouple the two parts of the resonance mode and obtain
\begin{equation}
\omega_R=\frac{\sqrt{K_{l}f(r_{0})}}{r_{0}}\sim \frac{l\sqrt{f(r_{0})}}{r_{0}},
\end{equation}
\begin{equation}
\omega_I=\frac{\sqrt{2}(2n+1)r_{0}\sqrt{-f(r_{0})V_{0}^{\prime\prime}}}{4\sqrt{K_{l}}}.
\end{equation}

We can define dimensionless quantities 
\begin{equation}
\bar{Q}\equiv\frac{\sqrt{d-2}}{\sqrt{2(d-3)}}\frac{Q}{M},\,\bar{L}\equiv\frac{L}{M^{\frac{1}{d-3}}},
\end{equation}
so that for $d=4$, the explicit expression of the imaginary part of the resonance frequency reads ($M=1$)
\begin{equation}
 \omega_I =\frac{2 \sqrt{2 y_4}}{(x_4+3)^3 \bar{L}},
\end{equation}
with
\begin{equation}\label{req1}
x_4=\sqrt{9-8 \bar{Q}^2},
\end{equation}
\begin{equation}
\begin{aligned}
y_4=&-12 (5 x+39) \bar{Q}^4+54 (5 x+21) \bar{Q}^2\\&+32 \bar{Q}^6-243 (x+3)\\&+\bar{L}^2 \left[-(7 x+33) \bar{Q}^2+8 \bar{Q}^4+9 (x+3)\right].
\end{aligned}
\end{equation}
For $d=5$, we have 
\begin{equation}
\omega_I=\frac{\pi  \bar{L}^2 \left(3 (x_5-3) \bar{Q}^2-4 x_5+8\right)+18 \bar{Q}^4}{3 \sqrt{6} (x_5-2) y_5 \bar{L}},
\end{equation}
with
\begin{equation}\label{req2}
x_5=\sqrt{4-3 \bar{Q}^2},
\end{equation}
\be
\bal
y_5=\sqrt{\frac{(x-2) \bar{Q}^6 \left(\pi  \bar{L}^2 \left(3 (x-3) \bar{Q}^2-4 x+8\right)+18 \bar{Q}^4\right)}{9 (x-8) \bar{Q}^4-96 (x-3) \bar{Q}^2+128 (x-2)}}.
\eal
\ee
And for $d=6$, we have 
\begin{equation}
\omega_I=\frac{\sqrt[6]{3} y_6 \left(8 \left(x_6-10\right) \bar{Q}^2-25 \left(x_6-5\right)\right)}{32 \left(x_6-5\right) \bar{L} \bar{Q}^2},
\end{equation}
with
\be
x_{6}=\sqrt{25-16 \bar{Q}^2},
\ee
\be
y_{6}=\sqrt{\frac{\left(x_6-5\right) \left(z_6-64\ 3^{2/3} \bar{Q}^2\right)}{8 \left(x_6-10\right) \bar{Q}^2-25 \left(x_6-5\right)}},
\ee
\be
z_{6}=\frac{3 \sqrt[3]{2} \pi ^{2/3} \left(5-x_6\right){}^{2/3} \bar{L}^2 \left(8 \bar{Q}^2+x_6-5\right)}{\bar{Q}^{4/3}}.
\ee

We have plotted the relaxation time of the RNdS black hole perturbed by the neutral massless scalar field in terms of the reduced electric charge of the black hole for the four-dimensional case in Fig. \ref{ps1911}, five-dimensional case in Fig. \ref{rt1816} and six-dimensional case in Fig. \ref{rt1618}. As we can see, for all cases, with the increasing of the electric charge of the black hole, the relaxation time first decreases and then increases. For the four-dimensional composed RNdS black hole/neutral massless scalar field system  we have the fastest relaxation rate at $\bar{Q}=0.726$, a value obtained in \cite{Hod:2018ifz}.

To further analyze the varying patterns of the relaxation rate, we take a look at the parameter space. In Fig. \ref{ps2311}, we have shown the parameter spaces of the four-dimensional, five-dimensional and six-dimensional RNdS black holes perturbed by the neutral massless scalar fields. We find that all the left diagrams in Figs. \ref{ps1911}, \ref{rt1816} and \ref{rt1618} correspond to the blue region in Fig. \ref{ps2311} and all the right diagrams in Figs. \ref{ps1911}, \ref{rt1816} and \ref{rt1618} correspond to the white region in Fig. \ref{ps2311}.

Basically, from Figs. \ref{ps1911}, \ref{rt1816} and \ref{rt1618}, it is not hard to see that, with the increasing of the cosmological constant, the fastest relaxation rate will increase, and the corresponding electric charge of the black hole will also increase. To see this more explicitly, we show Tables \ref{tbl:table1}, \ref{tbl:table2} and \ref{tbl:table3}. Interestingly, besides the aforementioned property of the relaxation rate, we also find another subtle property. For the four-dimensional case, we can see that when the cosmological constant goes to zero, the corresponding electric charge of the black hole maximizing the relaxation rate of the system decreases from $Q_{\text{ext}}$ to 0.726, a non-zero value. In contrast, for the higher-dimensional case, when the cosmological constant decreases, the electric charge that makes the relaxation rate fastest is decreases from maximum to zero. This is just in agreement with the result in \cite{Zhang:2019ynp}, where we found that the changing style of the composed higher-dimensional RN black hole/neutral massless scalar field system is different from the four-dimensional counterpart.

\section{Conclusion and Discussion}\label{chapter4}
We investigated the relaxation rate of the RNdS black hole perturbed by the neutral massless scalar field. To this end, we first showed the radial perturbation equation together with the physically reasonable boundary conditions. Then in the eikonal limit, we explicitly wrote down the quasinormal resonance frequency of the composed black hole/neutral massless scalar field system. We analyzed the relaxation rate of the system for four-dimensional case and higher-dimensional cases.

We found some common properties among different dimensional composed RNdS black hole/neutral massless scalar field systems. First, we got to know that the fastest relaxation rate (or the shortest relaxation time) increases with the increasing cosmological constant. Second, we showed that the critical value of the electric charge that makes the relaxation time shortest also increases with the increasing cosmological constant. 

We also found some differences between the four-dimensional composed system and its higher-dimensional counterpart. First, the structures of the parameter space are different, as shown in Fig. \ref{ps2311}. For the four-dimensional case, the largest value of the electric charge $Q_{\text{ext}}$ that the black hole can hold to ensure the existence of its three horizons coincides with the largest one that the black hole can take to ensure the existence of the system's quasinormal resonance frequency. For the higher-dimensional system, it is different. The occurrence of the red region in Fig. \ref{ps2311} originates from that the maximum value $Q_{\text{ext}}$ ensuring the existence of the QNM is smaller than the one ensuring the existence of the black hole's three horizons. Second, we found that, the extreme values of the electric charge of the black holes that make the relaxation time of the system shortest when the cosmological constant vanishes are different. For the four-dimensional system, when the cosmological constant decreases to zero from $Q_{\text{ext}}$, the fastest relaxation rate of the system decreases, and  the corresponding critical electric charge decreases to 0.726, which is non-zero; however, for the higher-dimensional system, when the cosmological constant decreases to be vanishing, the critical electric charge of the black hole that makes the relaxation rate of the system fastest decreases continuously and gradually to zero which is just the higher-dimensional Schwarzschild black hole limit. We verify our relaxation time data  by using the mathematical package in \cite{Jansen:2017oag}. We have checked our calculation by using the results in \cite{,Liu:2019lon}. As shown in Table \ref{tbl:table4}, the WKB method we used in the large $l$ eikonal regime is consistent with the numerical spectral method.

This is reminiscent of the results in \cite{Zhang:2019ynp}. From (\ref{effeff2}), it is not difficult to see that the relaxation time of the RNdS black hole perturbed by the neutral massless scalar field is related to the effective potential of the system by $\tau\sim V_{0}/V_{0}^{(2)}$. In the eikonal regime $l\gg 1$, the approximated effective potential  \eqref{appveff} for the composed RNdS black hole/neutral massless scalar field system is coincidentally  different from that for the composed RN black hole/neutral massless scalar field system only with a constant term related to the cosmological constant. Consequently, the patterns of $V_{0}$ and $V_{0}^{(2)}$ for the composed RN black hole/neutral massless scalar field system in \cite{Zhang:2019ynp} are similar to those for the composed RNdS black hole/neutral massless scalar field system. And this just explains why there are different dependent rules of the relaxation time on the charge between four-dimensional case and higher dimensional cases at the zero-cosmological constant limit. It also partly accounts for that the critical charge decreases with the decreasing cosmological constant, as the signs of the electric term and the cosmological constant term are contrary.

It was found in \cite{Zhu:2014sya,Brady:1996za} that, when $l=0$, following with some quasinormal oscillations, the perturbed neutral massless scalar field will eventually settle down to a constant value, which means that the RNdS black hole perturbed by the neutral massless scalar field is prone to instability. However, the perturbed field will fall off exponentially for the $l=1$ \cite{Zhu:2014sya}, $l=2$ \cite{Brady:1996za} and $l=3$ \cite{Molina:2003dc} modes, which means that the black hole is stable against the neutral massless scalar field. In the present work, we analytically calculated the quasinormal resonance modes in the eikonal regime $l\gg 1$ and our result is consistent with those former ones, as the relaxation time of the composed RNdS black hole/neutral massless scalar field system is finite for all allowed regions of the cosmological constant $\Lambda$, only softly varies with the electric charge of the black hole $Q$. The result is also consistent with the assertion which states that higher $l$ modes are more stable \cite{Konoplya:2020juj}. The RNdS black hole is found to be unstable at the $l=0$ mode once the perturbed field is charged, but this can be avoided by increasing the scalar field mass \cite{Konoplya:2014lha,Zhu:2014sya}. Thus, we suspect that the quality of the relaxation time for the composed RNdS black hole/massive charged black hole system may be closely  related to the angular index.

Our work extends former investigations of the relaxation rate of the asymptotically flat RN black hole perturbed by neutral massless scalar field to the asymptotically de Sitter case. There are two directions we may work on further. One is to study the characteristics of the asymptotically de Sitter rotating black hole perturbed by matter fields and the other is to investigate the asymptotically de Sitter black holes in modified gravitational theories. The RNAdS black hole is also valuable to consider. Nevertheless, the WKB method used above can be considered only by adding some restrictions. One may study the near extreme black hole and restrict the AdS radius as $M^2/L^2\ll K_l$, so that the effective potential can be greatly simplified.

%

\section*{Acknowledgements}
Zhen Zhong is supported by the National Natural Science Foundation of China (Grant Nos. 11675015 and 11775022). Jie Jiang is supported by the National Natural Science Foundation of China (Grant Nos. 11775022 and 11873044). Ming Zhang is supported by the Initial Research Foundation of Jiangxi Normal University with Grant No. 12020023.

\appendix
\section{Extreme value of electric charge for four-dimensional RNdS black hole}\label{ap}
The blackening factor of the four-dimensional RNdS black hole can be expressed as (we set $M=1$)
\begin{equation}
\begin{aligned}
f(r)=&\frac{L^2 Q^2+L^2 r^2-2 L^2 r-r^4}{L^2 r^2}\\=&\frac{(r_c-r)(r-r_+)(r-r_-)(r-r_o)}{L^2 r^2},
\end{aligned}
\end{equation}
where
\begin{equation}
-(r_c+r_-+r_+)=r_m<0<r_-<r_+<r_c.
\end{equation}
We have the coefficients of numerator of the polynomial $f(r)$ as
\begin{equation}
a=-1,\,c=L^2,\,d=-2L^2,\,e=L^2 Q^2,
\end{equation}
then we can define 
\begin{equation}
\mathcal{D}=-8ac,\,\mathcal{E}=-8a^2 d,\,\mathcal{F}=16a^2c^2-64a^3e
\end{equation}
and
\begin{equation}\begin{aligned}
\mathcal{A}=\mathcal{D}^2-3\mathcal{F},\, \mathcal{B}=\mathcal{D}\mathcal{F}-9\mathcal{E}^2,\, \mathcal{C}=\mathcal{F}^2-3\mathcal{D}\mathcal{E}^2.
\end{aligned}\end{equation}
As
\begin{equation}
\mathcal{D}=8L^2>0,\,\mathcal{F}=16L^4+64L^2Q^2>0,
\end{equation}
We have four different real roots for $f(r)=0$ iff
\begin{equation}\begin{aligned}
\Delta&=\mathcal{B}^2-4\mathcal{A}\mathcal{C}\\&=L^4 \left(Q^2-1\right)+L^2 \left(8 Q^4-36 Q^2+27\right)+16 Q^6\\&<0,\label{2519}
\end{aligned}\end{equation}
which gives
\begin{equation}
Q_\text{ext}=\sqrt{\frac{L^4-2 L^2 z^{1/3}+108 L^2+z^{2/3}}{12 z^{1/3}}}=\bar{Q}_\text{ext},
\end{equation}
where
\begin{equation}
z=L^6-270 L^4-1458 L^2+6 \sqrt{3} \sqrt{-L^4 \left(2 L^2-27\right)^3}.
\end{equation}
From (\ref{2519}), to ensure the existence of $L$,  we should let $Q<3/(2 \sqrt{2})=1.06066\equiv Q_\text{ext}$, then we have $L>3\sqrt{3/2}=3.67423\equiv L_\text{ext}$. The Cauchy horizon coincides with the event horizon for $Q=Q_\text{ext}$, and there will be at most one positive radius for the quadratic equation $f(r)=0$ if $L< L_\text{ext}$.


\begin{thebibliography}{33}%
\makeatletter
\providecommand \@ifxundefined [1]{%
 \@ifx{#1\undefined}
}%
\providecommand \@ifnum [1]{%
 \ifnum #1\expandafter \@firstoftwo
 \else \expandafter \@secondoftwo
 \fi
}%
\providecommand \@ifx [1]{%
 \ifx #1\expandafter \@firstoftwo
 \else \expandafter \@secondoftwo
 \fi
}%
\providecommand \natexlab [1]{#1}%
\providecommand \enquote  [1]{``#1''}%
\providecommand \bibnamefont  [1]{#1}%
\providecommand \bibfnamefont [1]{#1}%
\providecommand \citenamefont [1]{#1}%
\providecommand \href@noop [0]{\@secondoftwo}%
\providecommand \href [0]{\begingroup \@sanitize@url \@href}%
\providecommand \@href[1]{\@@startlink{#1}\@@href}%
\providecommand \@@href[1]{\endgroup#1\@@endlink}%
\providecommand \@sanitize@url [0]{\catcode `\\12\catcode `\$12\catcode
  `\&12\catcode `\#12\catcode `\^12\catcode `\_12\catcode `\%12\relax}%
\providecommand \@@startlink[1]{}%
\providecommand \@@endlink[0]{}%
\providecommand \url  [0]{\begingroup\@sanitize@url \@url }%
\providecommand \@url [1]{\endgroup\@href {#1}{\urlprefix }}%
\providecommand \urlprefix  [0]{URL }%
\providecommand \Eprint [0]{\href }%
\providecommand \doibase [0]{http://dx.doi.org/}%
\providecommand \selectlanguage [0]{\@gobble}%
\providecommand \bibinfo  [0]{\@secondoftwo}%
\providecommand \bibfield  [0]{\@secondoftwo}%
\providecommand \translation [1]{[#1]}%
\providecommand \BibitemOpen [0]{}%
\providecommand \bibitemStop [0]{}%
\providecommand \bibitemNoStop [0]{.\EOS\space}%
\providecommand \EOS [0]{\spacefactor3000\relax}%
\providecommand \BibitemShut  [1]{\csname bibitem#1\endcsname}%
\let\auto@bib@innerbib\@empty
\bibitem [{\citenamefont {Berti}\ \emph {et~al.}(2009)\citenamefont {Berti},
  \citenamefont {Cardoso},\ and\ \citenamefont {Starinets}}]{Berti:2009kk}%
  \BibitemOpen
  \bibfield  {author} {\bibinfo {author} {\bibfnamefont {E.}~\bibnamefont
  {Berti}}, \bibinfo {author} {\bibfnamefont {V.}~\bibnamefont {Cardoso}}, \
  and\ \bibinfo {author} {\bibfnamefont {A.~O.}\ \bibnamefont {Starinets}},\
  }\href {\doibase 10.1088/0264-9381/26/16/163001} {\bibfield  {journal}
  {\bibinfo  {journal} {Class. Quant. Grav.}\ }\textbf {\bibinfo {volume}
  {26}},\ \bibinfo {pages} {163001} (\bibinfo {year} {2009})},\ \Eprint
  {http://arxiv.org/abs/0905.2975} {arXiv:0905.2975 [gr-qc]} \BibitemShut
  {NoStop}%
\bibitem [{\citenamefont {Cardoso}\ \emph {et~al.}(2018)\citenamefont
  {Cardoso}, \citenamefont {Costa}, \citenamefont {Destounis}, \citenamefont
  {Hintz},\ and\ \citenamefont {Jansen}}]{Cardoso:2017soq}%
  \BibitemOpen
  \bibfield  {author} {\bibinfo {author} {\bibfnamefont {V.}~\bibnamefont
  {Cardoso}}, \bibinfo {author} {\bibfnamefont {J.~L.}\ \bibnamefont {Costa}},
  \bibinfo {author} {\bibfnamefont {K.}~\bibnamefont {Destounis}}, \bibinfo
  {author} {\bibfnamefont {P.}~\bibnamefont {Hintz}}, \ and\ \bibinfo {author}
  {\bibfnamefont {A.}~\bibnamefont {Jansen}},\ }\href {\doibase
  10.1103/PhysRevLett.120.031103} {\bibfield  {journal} {\bibinfo  {journal}
  {Phys. Rev. Lett.}\ }\textbf {\bibinfo {volume} {120}},\ \bibinfo {pages}
  {031103} (\bibinfo {year} {2018})},\ \Eprint
  {http://arxiv.org/abs/1711.10502} {arXiv:1711.10502 [gr-qc]} \BibitemShut
  {NoStop}%
\bibitem [{\citenamefont {Hod}(2019)}]{Hod:2018dpx}%
  \BibitemOpen
  \bibfield  {author} {\bibinfo {author} {\bibfnamefont {S.}~\bibnamefont
  {Hod}},\ }\href {\doibase 10.1016/j.nuclphysb.2019.03.003} {\bibfield
  {journal} {\bibinfo  {journal} {Nucl. Phys.}\ }\textbf {\bibinfo {volume}
  {B941}},\ \bibinfo {pages} {636} (\bibinfo {year} {2019})},\ \Eprint
  {http://arxiv.org/abs/1801.07261} {arXiv:1801.07261 [gr-qc]} \BibitemShut
  {NoStop}%
\bibitem [{\citenamefont {Hod}(2015{\natexlab{a}})}]{Hod:2015hza}%
  \BibitemOpen
  \bibfield  {author} {\bibinfo {author} {\bibfnamefont {S.}~\bibnamefont
  {Hod}},\ }\href {\doibase 10.1103/PhysRevD.91.044047} {\bibfield  {journal}
  {\bibinfo  {journal} {Phys. Rev.}\ }\textbf {\bibinfo {volume} {D91}},\
  \bibinfo {pages} {044047} (\bibinfo {year} {2015}{\natexlab{a}})},\ \Eprint
  {http://arxiv.org/abs/1504.00009} {arXiv:1504.00009 [gr-qc]} \BibitemShut
  {NoStop}%
\bibitem [{\citenamefont {Hod}(2016{\natexlab{a}})}]{Hod:2016bas}%
  \BibitemOpen
  \bibfield  {author} {\bibinfo {author} {\bibfnamefont {S.}~\bibnamefont
  {Hod}},\ }\href {\doibase 10.1103/PhysRevD.94.044036} {\bibfield  {journal}
  {\bibinfo  {journal} {Phys. Rev.}\ }\textbf {\bibinfo {volume} {D94}},\
  \bibinfo {pages} {044036} (\bibinfo {year} {2016}{\natexlab{a}})},\ \Eprint
  {http://arxiv.org/abs/1609.07146} {arXiv:1609.07146 [gr-qc]} \BibitemShut
  {NoStop}%
\bibitem [{\citenamefont {Hod}(2016{\natexlab{b}})}]{Hod:2017www}%
  \BibitemOpen
  \bibfield  {author} {\bibinfo {author} {\bibfnamefont {S.}~\bibnamefont
  {Hod}},\ }\href {\doibase 10.1016/j.physletb.2016.10.069} {\bibfield
  {journal} {\bibinfo  {journal} {Phys. Lett.}\ }\textbf {\bibinfo {volume}
  {B763}},\ \bibinfo {pages} {275} (\bibinfo {year} {2016}{\natexlab{b}})},\
  \Eprint {http://arxiv.org/abs/1703.05333} {arXiv:1703.05333 [gr-qc]}
  \BibitemShut {NoStop}%
\bibitem [{\citenamefont {Hod}(2018{\natexlab{a}})}]{Hod:2018fet}%
  \BibitemOpen
  \bibfield  {author} {\bibinfo {author} {\bibfnamefont {S.}~\bibnamefont
  {Hod}},\ }\href {\doibase 10.1016/j.physletb.2018.09.039} {\bibfield
  {journal} {\bibinfo  {journal} {Phys. Lett.}\ }\textbf {\bibinfo {volume}
  {B786}},\ \bibinfo {pages} {217} (\bibinfo {year} {2018}{\natexlab{a}})},\
  \Eprint {http://arxiv.org/abs/1808.04077} {arXiv:1808.04077 [gr-qc]}
  \BibitemShut {NoStop}%
\bibitem [{\citenamefont {Hod}(2015{\natexlab{b}})}]{Hod:2017kpt}%
  \BibitemOpen
  \bibfield  {author} {\bibinfo {author} {\bibfnamefont {S.}~\bibnamefont
  {Hod}},\ }\href {\doibase 10.1016/j.physletb.2015.10.039} {\bibfield
  {journal} {\bibinfo  {journal} {Phys. Lett.}\ }\textbf {\bibinfo {volume}
  {B751}},\ \bibinfo {pages} {177} (\bibinfo {year} {2015}{\natexlab{b}})},\
  \Eprint {http://arxiv.org/abs/1707.06246} {arXiv:1707.06246 [gr-qc]}
  \BibitemShut {NoStop}%
\bibitem [{\citenamefont {Huang}\ \emph {et~al.}(2017)\citenamefont {Huang},
  \citenamefont {Liu}, \citenamefont {Zhai},\ and\ \citenamefont
  {Li}}]{Huang:2017whw}%
  \BibitemOpen
  \bibfield  {author} {\bibinfo {author} {\bibfnamefont {Y.}~\bibnamefont
  {Huang}}, \bibinfo {author} {\bibfnamefont {D.-J.}\ \bibnamefont {Liu}},
  \bibinfo {author} {\bibfnamefont {X.-H.}\ \bibnamefont {Zhai}}, \ and\
  \bibinfo {author} {\bibfnamefont {X.-Z.}\ \bibnamefont {Li}},\ }\href
  {\doibase 10.1088/1361-6382/aa7964} {\bibfield  {journal} {\bibinfo
  {journal} {Class. Quant. Grav.}\ }\textbf {\bibinfo {volume} {34}},\ \bibinfo
  {pages} {155002} (\bibinfo {year} {2017})},\ \Eprint
  {http://arxiv.org/abs/1706.04441} {arXiv:1706.04441 [gr-qc]} \BibitemShut
  {NoStop}%
\bibitem [{\citenamefont {Li}\ \emph {et~al.}(2019)\citenamefont {Li},
  \citenamefont {Zhao}, \citenamefont {Zi},\ and\ \citenamefont
  {Chen}}]{Li:2019tns}%
  \BibitemOpen
  \bibfield  {author} {\bibinfo {author} {\bibfnamefont {R.}~\bibnamefont
  {Li}}, \bibinfo {author} {\bibfnamefont {Y.}~\bibnamefont {Zhao}}, \bibinfo
  {author} {\bibfnamefont {T.}~\bibnamefont {Zi}}, \ and\ \bibinfo {author}
  {\bibfnamefont {X.}~\bibnamefont {Chen}},\ }\href {\doibase
  10.1103/PhysRevD.99.084045} {\bibfield  {journal} {\bibinfo  {journal} {Phys.
  Rev.}\ }\textbf {\bibinfo {volume} {D99}},\ \bibinfo {pages} {084045}
  (\bibinfo {year} {2019})}\BibitemShut {NoStop}%
\bibitem [{\citenamefont {Nollert}(1999)}]{Nollert:1999ji}%
  \BibitemOpen
  \bibfield  {author} {\bibinfo {author} {\bibfnamefont {H.-P.}\ \bibnamefont
  {Nollert}},\ }\href {\doibase 10.1088/0264-9381/16/12/201} {\bibfield
  {journal} {\bibinfo  {journal} {Class. Quant. Grav.}\ }\textbf {\bibinfo
  {volume} {16}},\ \bibinfo {pages} {R159} (\bibinfo {year}
  {1999})}\BibitemShut {NoStop}%
\bibitem [{\citenamefont {Cardoso}\ and\ \citenamefont
  {Lemos}(2001)}]{Cardoso:2001hn}%
  \BibitemOpen
  \bibfield  {author} {\bibinfo {author} {\bibfnamefont {V.}~\bibnamefont
  {Cardoso}}\ and\ \bibinfo {author} {\bibfnamefont {J.~P.~S.}\ \bibnamefont
  {Lemos}},\ }\href {\doibase 10.1103/PhysRevD.63.124015} {\bibfield  {journal}
  {\bibinfo  {journal} {Phys. Rev.}\ }\textbf {\bibinfo {volume} {D63}},\
  \bibinfo {pages} {124015} (\bibinfo {year} {2001})},\ \Eprint
  {http://arxiv.org/abs/gr-qc/0101052} {arXiv:gr-qc/0101052 [gr-qc]}
  \BibitemShut {NoStop}%
\bibitem [{\citenamefont {Hod}(1998)}]{Hod:1998vk}%
  \BibitemOpen
  \bibfield  {author} {\bibinfo {author} {\bibfnamefont {S.}~\bibnamefont
  {Hod}},\ }\href {\doibase 10.1103/PhysRevLett.81.4293} {\bibfield  {journal}
  {\bibinfo  {journal} {Phys. Rev. Lett.}\ }\textbf {\bibinfo {volume} {81}},\
  \bibinfo {pages} {4293} (\bibinfo {year} {1998})},\ \Eprint
  {http://arxiv.org/abs/gr-qc/9812002} {arXiv:gr-qc/9812002 [gr-qc]}
  \BibitemShut {NoStop}%
\bibitem [{\citenamefont {Carter}(1971)}]{Carter:1971zc}%
  \BibitemOpen
  \bibfield  {author} {\bibinfo {author} {\bibfnamefont {B.}~\bibnamefont
  {Carter}},\ }\href {\doibase 10.1103/PhysRevLett.26.331} {\bibfield
  {journal} {\bibinfo  {journal} {Phys. Rev. Lett.}\ }\textbf {\bibinfo
  {volume} {26}},\ \bibinfo {pages} {331} (\bibinfo {year} {1971})}\BibitemShut
  {NoStop}%
\bibitem [{\citenamefont {Hawking}(1972)}]{Hawking:1971vc}%
  \BibitemOpen
  \bibfield  {author} {\bibinfo {author} {\bibfnamefont {S.~W.}\ \bibnamefont
  {Hawking}},\ }\href {\doibase 10.1007/BF01877517} {\bibfield  {journal}
  {\bibinfo  {journal} {Commun. Math. Phys.}\ }\textbf {\bibinfo {volume}
  {25}},\ \bibinfo {pages} {152} (\bibinfo {year} {1972})}\BibitemShut
  {NoStop}%
\bibitem [{\citenamefont {Robinson}(1975)}]{Robinson:1975bv}%
  \BibitemOpen
  \bibfield  {author} {\bibinfo {author} {\bibfnamefont {D.~C.}\ \bibnamefont
  {Robinson}},\ }\href {\doibase 10.1103/PhysRevLett.34.905} {\bibfield
  {journal} {\bibinfo  {journal} {Phys. Rev. Lett.}\ }\textbf {\bibinfo
  {volume} {34}},\ \bibinfo {pages} {905} (\bibinfo {year} {1975})}\BibitemShut
  {NoStop}%
\bibitem [{\citenamefont {Ruffini}\ and\ \citenamefont
  {Wheeler}(1971)}]{Ruffini:1971bza}%
  \BibitemOpen
  \bibfield  {author} {\bibinfo {author} {\bibfnamefont {R.}~\bibnamefont
  {Ruffini}}\ and\ \bibinfo {author} {\bibfnamefont {J.~A.}\ \bibnamefont
  {Wheeler}},\ }\href {\doibase 10.1063/1.3022513} {\bibfield  {journal}
  {\bibinfo  {journal} {Phys. Today}\ }\textbf {\bibinfo {volume} {24}},\
  \bibinfo {pages} {30} (\bibinfo {year} {1971})}\BibitemShut {NoStop}%
\bibitem [{\citenamefont {Zhu}\ \emph {et~al.}(2014)\citenamefont {Zhu},
  \citenamefont {Zhang}, \citenamefont {Pellicer}, \citenamefont {Wang},\ and\
  \citenamefont {Abdalla}}]{Zhu:2014sya}%
  \BibitemOpen
  \bibfield  {author} {\bibinfo {author} {\bibfnamefont {Z.}~\bibnamefont
  {Zhu}}, \bibinfo {author} {\bibfnamefont {S.-J.}\ \bibnamefont {Zhang}},
  \bibinfo {author} {\bibfnamefont {C.~E.}\ \bibnamefont {Pellicer}}, \bibinfo
  {author} {\bibfnamefont {B.}~\bibnamefont {Wang}}, \ and\ \bibinfo {author}
  {\bibfnamefont {E.}~\bibnamefont {Abdalla}},\ }\href {\doibase
  10.1103/PhysRevD.90.044042, 10.1103/PhysRevD.90.049904} {\bibfield  {journal}
  {\bibinfo  {journal} {Phys. Rev.}\ }\textbf {\bibinfo {volume} {D90}},\
  \bibinfo {pages} {044042} (\bibinfo {year} {2014})},\ \bibinfo {note}
  {[Addendum: Phys. Rev.D90,no.4,049904(2014)]},\ \Eprint
  {http://arxiv.org/abs/1405.4931} {arXiv:1405.4931 [hep-th]} \BibitemShut
  {NoStop}%
\bibitem [{\citenamefont {Konoplya}\ and\ \citenamefont
  {Zhidenko}(2014)}]{Konoplya:2014lha}%
  \BibitemOpen
  \bibfield  {author} {\bibinfo {author} {\bibfnamefont {R.~A.}\ \bibnamefont
  {Konoplya}}\ and\ \bibinfo {author} {\bibfnamefont {A.}~\bibnamefont
  {Zhidenko}},\ }\href {\doibase 10.1103/PhysRevD.90.064048} {\bibfield
  {journal} {\bibinfo  {journal} {Phys. Rev.}\ }\textbf {\bibinfo {volume}
  {D90}},\ \bibinfo {pages} {064048} (\bibinfo {year} {2014})},\ \Eprint
  {http://arxiv.org/abs/1406.0019} {arXiv:1406.0019 [hep-th]} \BibitemShut
  {NoStop}%
\bibitem [{\citenamefont {Hod}(2016{\natexlab{c}})}]{Hod:2016jqt}%
  \BibitemOpen
  \bibfield  {author} {\bibinfo {author} {\bibfnamefont {S.}~\bibnamefont
  {Hod}},\ }\href {\doibase 10.1016/j.physletb.2016.08.008} {\bibfield
  {journal} {\bibinfo  {journal} {Phys. Lett.}\ }\textbf {\bibinfo {volume}
  {B761}},\ \bibinfo {pages} {53} (\bibinfo {year} {2016}{\natexlab{c}})},\
  \Eprint {http://arxiv.org/abs/1609.01297} {arXiv:1609.01297 [gr-qc]}
  \BibitemShut {NoStop}%
\bibitem [{\citenamefont {Zhang}\ \emph
  {et~al.}(2019{\natexlab{a}})\citenamefont {Zhang}, \citenamefont {Jiang},\
  and\ \citenamefont {Zhong}}]{Zhang:2018jgj}%
  \BibitemOpen
  \bibfield  {author} {\bibinfo {author} {\bibfnamefont {M.}~\bibnamefont
  {Zhang}}, \bibinfo {author} {\bibfnamefont {J.}~\bibnamefont {Jiang}}, \ and\
  \bibinfo {author} {\bibfnamefont {Z.}~\bibnamefont {Zhong}},\ }\href
  {\doibase 10.1016/j.physletb.2018.10.072} {\bibfield  {journal} {\bibinfo
  {journal} {Phys. Lett.}\ }\textbf {\bibinfo {volume} {B789}},\ \bibinfo
  {pages} {13} (\bibinfo {year} {2019}{\natexlab{a}})},\ \Eprint
  {http://arxiv.org/abs/1811.04183} {arXiv:1811.04183 [gr-qc]} \BibitemShut
  {NoStop}%
\bibitem [{\citenamefont {Hod}(2018{\natexlab{b}})}]{Hod:2018ifz}%
  \BibitemOpen
  \bibfield  {author} {\bibinfo {author} {\bibfnamefont {S.}~\bibnamefont
  {Hod}},\ }\href {\doibase 10.1140/epjc/s10052-018-6422-8} {\bibfield
  {journal} {\bibinfo  {journal} {Eur. Phys. J.}\ }\textbf {\bibinfo {volume}
  {C78}},\ \bibinfo {pages} {935} (\bibinfo {year} {2018}{\natexlab{b}})},\
  \Eprint {http://arxiv.org/abs/1812.01014} {arXiv:1812.01014 [gr-qc]}
  \BibitemShut {NoStop}%
\bibitem [{\citenamefont {Zhang}\ \emph
  {et~al.}(2019{\natexlab{b}})\citenamefont {Zhang}, \citenamefont {Jiang},\
  and\ \citenamefont {Zhong}}]{Zhang:2019ynp}%
  \BibitemOpen
  \bibfield  {author} {\bibinfo {author} {\bibfnamefont {M.}~\bibnamefont
  {Zhang}}, \bibinfo {author} {\bibfnamefont {J.}~\bibnamefont {Jiang}}, \ and\
  \bibinfo {author} {\bibfnamefont {Z.}~\bibnamefont {Zhong}},\ }\href
  {\doibase 10.1016/j.physletb.2019.134959} {\bibfield  {journal} {\bibinfo
  {journal} {Phys. Lett.}\ }\textbf {\bibinfo {volume} {B798}},\ \bibinfo
  {pages} {134959} (\bibinfo {year} {2019}{\natexlab{b}})},\ \Eprint
  {http://arxiv.org/abs/1909.08562} {arXiv:1909.08562 [gr-qc]} \BibitemShut
  {NoStop}%
\bibitem [{\citenamefont {Pani}\ \emph
  {et~al.}(2013{\natexlab{a}})\citenamefont {Pani}, \citenamefont {Berti},\
  and\ \citenamefont {Gualtieri}}]{Pani:2013wsa}%
  \BibitemOpen
  \bibfield  {author} {\bibinfo {author} {\bibfnamefont {P.}~\bibnamefont
  {Pani}}, \bibinfo {author} {\bibfnamefont {E.}~\bibnamefont {Berti}}, \ and\
  \bibinfo {author} {\bibfnamefont {L.}~\bibnamefont {Gualtieri}},\ }\href
  {\doibase 10.1103/PhysRevD.88.064048} {\bibfield  {journal} {\bibinfo
  {journal} {Phys. Rev.}\ }\textbf {\bibinfo {volume} {D88}},\ \bibinfo {pages}
  {064048} (\bibinfo {year} {2013}{\natexlab{a}})},\ \Eprint
  {http://arxiv.org/abs/1307.7315} {arXiv:1307.7315 [gr-qc]} \BibitemShut
  {NoStop}%
\bibitem [{\citenamefont {Pani}\ \emph
  {et~al.}(2013{\natexlab{b}})\citenamefont {Pani}, \citenamefont {Berti},\
  and\ \citenamefont {Gualtieri}}]{Pani:2013ija}%
  \BibitemOpen
  \bibfield  {author} {\bibinfo {author} {\bibfnamefont {P.}~\bibnamefont
  {Pani}}, \bibinfo {author} {\bibfnamefont {E.}~\bibnamefont {Berti}}, \ and\
  \bibinfo {author} {\bibfnamefont {L.}~\bibnamefont {Gualtieri}},\ }\href
  {\doibase 10.1103/PhysRevLett.110.241103} {\bibfield  {journal} {\bibinfo
  {journal} {Phys. Rev. Lett.}\ }\textbf {\bibinfo {volume} {110}},\ \bibinfo
  {pages} {241103} (\bibinfo {year} {2013}{\natexlab{b}})},\ \Eprint
  {http://arxiv.org/abs/1304.1160} {arXiv:1304.1160 [gr-qc]} \BibitemShut
  {NoStop}%
\bibitem [{\citenamefont {Konoplya}(2003)}]{Konoplya:2003ii}%
  \BibitemOpen
  \bibfield  {author} {\bibinfo {author} {\bibfnamefont {R.~A.}\ \bibnamefont
  {Konoplya}},\ }\href {\doibase 10.1103/PhysRevD.68.024018} {\bibfield
  {journal} {\bibinfo  {journal} {Phys. Rev.}\ }\textbf {\bibinfo {volume}
  {D68}},\ \bibinfo {pages} {024018} (\bibinfo {year} {2003})},\ \Eprint
  {http://arxiv.org/abs/gr-qc/0303052} {arXiv:gr-qc/0303052 [gr-qc]}
  \BibitemShut {NoStop}%
\bibitem [{\citenamefont {Iyer}\ and\ \citenamefont
  {Will}(1987)}]{Iyer:1986np}%
  \BibitemOpen
  \bibfield  {author} {\bibinfo {author} {\bibfnamefont {S.}~\bibnamefont
  {Iyer}}\ and\ \bibinfo {author} {\bibfnamefont {C.~M.}\ \bibnamefont
  {Will}},\ }\href {\doibase 10.1103/PhysRevD.35.3621} {\bibfield  {journal}
  {\bibinfo  {journal} {Phys. Rev.}\ }\textbf {\bibinfo {volume} {D35}},\
  \bibinfo {pages} {3621} (\bibinfo {year} {1987})}\BibitemShut {NoStop}%
\bibitem [{\citenamefont {Schutz}\ and\ \citenamefont
  {Will}(1985)}]{Schutz:1985zz}%
  \BibitemOpen
  \bibfield  {author} {\bibinfo {author} {\bibfnamefont {B.~F.}\ \bibnamefont
  {Schutz}}\ and\ \bibinfo {author} {\bibfnamefont {C.~M.}\ \bibnamefont
  {Will}},\ }\href {\doibase 10.1086/184453} {\bibfield  {journal} {\bibinfo
  {journal} {Astrophys. J.}\ }\textbf {\bibinfo {volume} {291}},\ \bibinfo
  {pages} {L33} (\bibinfo {year} {1985})}\BibitemShut {NoStop}%
\bibitem [{\citenamefont {Jansen}(2017)}]{Jansen:2017oag}%
  \BibitemOpen
  \bibfield  {author} {\bibinfo {author} {\bibfnamefont {A.}~\bibnamefont
  {Jansen}},\ }\href {\doibase 10.1140/epjp/i2017-11825-9} {\bibfield
  {journal} {\bibinfo  {journal} {Eur. Phys. J. Plus}\ }\textbf {\bibinfo
  {volume} {132}},\ \bibinfo {pages} {546} (\bibinfo {year} {2017})},\ \Eprint
  {http://arxiv.org/abs/1709.09178} {arXiv:1709.09178 [gr-qc]} \BibitemShut
  {NoStop}%
\bibitem [{\citenamefont {Liu}\ \emph {et~al.}(2019)\citenamefont {Liu},
  \citenamefont {Tang}, \citenamefont {Destounis}, \citenamefont {Wang},
  \citenamefont {Papantonopoulos},\ and\ \citenamefont {Zhang}}]{Liu:2019lon}%
  \BibitemOpen
  \bibfield  {author} {\bibinfo {author} {\bibfnamefont {H.}~\bibnamefont
  {Liu}}, \bibinfo {author} {\bibfnamefont {Z.}~\bibnamefont {Tang}}, \bibinfo
  {author} {\bibfnamefont {K.}~\bibnamefont {Destounis}}, \bibinfo {author}
  {\bibfnamefont {B.}~\bibnamefont {Wang}}, \bibinfo {author} {\bibfnamefont
  {E.}~\bibnamefont {Papantonopoulos}}, \ and\ \bibinfo {author} {\bibfnamefont
  {H.}~\bibnamefont {Zhang}},\ }\href {\doibase 10.1007/JHEP03(2019)187}
  {\bibfield  {journal} {\bibinfo  {journal} {JHEP}\ }\textbf {\bibinfo
  {volume} {03}},\ \bibinfo {pages} {187} (\bibinfo {year} {2019})},\ \Eprint
  {http://arxiv.org/abs/1902.01865} {arXiv:1902.01865 [gr-qc]} \BibitemShut
  {NoStop}%
\bibitem [{\citenamefont {Brady}\ \emph {et~al.}(1997)\citenamefont {Brady},
  \citenamefont {Chambers}, \citenamefont {Krivan},\ and\ \citenamefont
  {Laguna}}]{Brady:1996za}%
  \BibitemOpen
  \bibfield  {author} {\bibinfo {author} {\bibfnamefont {P.~R.}\ \bibnamefont
  {Brady}}, \bibinfo {author} {\bibfnamefont {C.~M.}\ \bibnamefont {Chambers}},
  \bibinfo {author} {\bibfnamefont {W.}~\bibnamefont {Krivan}}, \ and\ \bibinfo
  {author} {\bibfnamefont {P.}~\bibnamefont {Laguna}},\ }\href {\doibase
  10.1103/PhysRevD.55.7538} {\bibfield  {journal} {\bibinfo  {journal} {Phys.
  Rev.}\ }\textbf {\bibinfo {volume} {D55}},\ \bibinfo {pages} {7538} (\bibinfo
  {year} {1997})},\ \Eprint {http://arxiv.org/abs/gr-qc/9611056}
  {arXiv:gr-qc/9611056 [gr-qc]} \BibitemShut {NoStop}%
\bibitem [{\citenamefont {Molina}\ \emph {et~al.}(2004)\citenamefont {Molina},
  \citenamefont {Giugno}, \citenamefont {Abdalla},\ and\ \citenamefont
  {Saa}}]{Molina:2003dc}%
  \BibitemOpen
  \bibfield  {author} {\bibinfo {author} {\bibfnamefont {C.}~\bibnamefont
  {Molina}}, \bibinfo {author} {\bibfnamefont {D.}~\bibnamefont {Giugno}},
  \bibinfo {author} {\bibfnamefont {E.}~\bibnamefont {Abdalla}}, \ and\
  \bibinfo {author} {\bibfnamefont {A.}~\bibnamefont {Saa}},\ }\href {\doibase
  10.1103/PhysRevD.69.104013} {\bibfield  {journal} {\bibinfo  {journal} {Phys.
  Rev.}\ }\textbf {\bibinfo {volume} {D69}},\ \bibinfo {pages} {104013}
  (\bibinfo {year} {2004})},\ \Eprint {http://arxiv.org/abs/gr-qc/0309079}
  {arXiv:gr-qc/0309079 [gr-qc]} \BibitemShut {NoStop}%
\bibitem [{\citenamefont {Konoplya}\ and\ \citenamefont
  {Zhidenko}(2020)}]{Konoplya:2020juj}%
  \BibitemOpen
  \bibfield  {author} {\bibinfo {author} {\bibfnamefont {R.}~\bibnamefont
  {Konoplya}}\ and\ \bibinfo {author} {\bibfnamefont {A.}~\bibnamefont
  {Zhidenko}},\ }\href@noop {} {\  (\bibinfo {year} {2020})},\ \Eprint
  {http://arxiv.org/abs/2003.12492} {arXiv:2003.12492 [gr-qc]} \BibitemShut
  {NoStop}%
\end{thebibliography}
%

\end{document}